\documentstyle[aps,epsf,twocolumn]{revtex}
\tighten
\begin{document}
\draft
\twocolumn[\hsize\textwidth\columnwidth\hsize\csname
@twocolumnfalse\endcsname


\title{Mode sum regularization approach for the self-force in black
hole spacetime}
\author{Leor Barack and Amos Ori}
\address {Department of Physics,
          Technion---Israel Institute of Technology, Haifa, 32000, Israel}
\date{\today}
\maketitle


\begin{abstract}

We present a method for calculating the self-force (the ``radiation reaction
force'') acting on a charged particle moving in a strong field orbit in
black hole spacetime.
In this approach, one first calculates the contribution to the self-force
due to each multipole mode of the particle's field. Then, the sum
over modes is evaluated, subject to a certain regularization procedure.
Here we develop this regularization procedure for a scalar charge on a
Schwarzschild background, and present the results of its implementation
for radial trajectories (not necessarily geodesic).

\end{abstract}
\pacs{04.25-g, 04.30.Db, 04.70.Bw}

\vspace{3ex} ]

A promising source for the LISA space-based gravitational wave detector
is the capture of a small compact object by a supermassive black hole.
The waveforms in
this scenario are expected to be well approximated by a model of a small
pointlike mass (a ``particle'') moving near the black hole\cite{Schutz}.
One also expects such a model to provide valuable information in more
general scenarios involving strongly gravitating two-body systems.
To accurately predict the orbital evolution of a particle near a black hole
(thus allowing the design of templates needed for detection and interpretation
of gravitational wave forms), it is necessary to include nongeodesic effects
caused by the interaction of the particle with its own field. Namely, one must
tackle the problem of calculating the {\it self-force} (sometimes dubbed
the ``radiation reaction'' force) experienced by a particle in a curved
spacetime.
The usual approach to this problem is via a perturbation analysis, in which
the charge $q$ of the particle (which may represent its mass or
its electric charge, or---in the toy model considered here---its scalar
charge) is assumed to be much smaller than the mass
$M$ of the black hole, and one looks for the leading order
($\propto q^2$) self-force effect.

There has been extensive effort, particularly over the last few years,
to develop methods for calculating the self-force in curved backgrounds.
The main challenge involved in this calculation is to deal with
the various infinite quantities associated with the divergence of the
field at the particle's location, which necessitates the introduction
of an appropriate regularization technique.
One calculation method, pioneered
long ago by DeWitt and Brehme\cite{DeWitt}, deduces the particle's equation
of motion by imposing local energy-momentum conservation on a world-tube
surrounding the particle's worldline\cite{Mino}.
Another approach\cite{balance} is based on computing the flux to infinity
(and across the horizon) of conserved quantities which are
tightly related to the orbit's constants of motion,
particularly the particle's energy $E$ and azimuthal angular momentum $L_z$
in the Schwarzschild and Kerr cases.
This approach, however, is inadequate for analyzing the evolution of the
third constant of motion in Kerr, the Carter constant $Q$.
Also, such techniques do not account for the
non-dissipative part of the self-force\cite{Wiseman,Burko3}.
So far, calculations of the self-force in actual examples have been
restricted to very few simple cases, such as static charges in the
Schwarzschild and Reissner-Nordstr\"{o}m geometries\cite{SW,Wiseman}.
A general formal framework for obtaining equations of motion for a test
particle in a curved spacetime has been formulated recently by Quinn and
Wald\cite{QW1,QW2};
however, the practical implementation of this approach
in actual calculations remained, so far, a challenging task.

Previously, Ori proposed\cite{Ori} that radiation reaction effects
may be calculated by evaluating the contribution of each
Fourier-multipole mode $\ell m\omega$ of the retarded field to the radiative
evolution, through the local force applied on the particle by this mode,
and then summing over all modes.
This approach has two advantages:
First, each individual mode of the field turns out to be continuous (and the
corresponding self-force to be bounded) even at the particle's location.
Secondly, calculating each $\ell m\omega$ mode of the field becomes a
relatively simple task, as it only requires the solution of an
ordinary differential equation (DE).
In \cite{Ori} this sum-over-modes approach has been proposed
for the calculation of the adiabatic,
orbit-integrated, evolution rate of the three
constants of motion in Kerr, i.e., $E$, $L_z$, and $Q$.\footnote
{The mode sum for the adiabatic, orbit-integrated, evolution
rate of $E$, $L_z$ was shown to converge\cite{Ori}.
It is not clear yet whether the corresponding
mode sum for $Q$ converges or not.}
It might have been hoped that the same
method would also be useful for calculating the momentary self-force;
however, this naive procedure for calculating the self-force turns
out inapplicable: Although each mode yields a finite contribution,
the sum over all modes is found to diverge. This is the situation even in
the simple case of a static scalar charge in flat space: For such a
charge, located at a distance $r_0$ from the origin of coordinates
(with respect to which the spherical harmonics are defined), the
contribution of each multipole $\ell$ to the radial component of the
self-force is the same: $-q^2/(2r_0^2)$.
Obviously, the sum over $\ell$ diverges.
To overcome this type of divergence,
one must introduce a certain regularization procedure into
the calculation.

In this paper we describe the basic elements of a new regularization scheme
for the mode sum.
As this method does not involve any weak-field or slow-motion
approximations, it should allow effective calculations of the
self-force for strong field orbits.
Here, we outline the method for a scalar particle moving
on a Schwarzschild background, and present the final results for
the special case of radial motion (not necessarily geodesic).
We also mention some results, derived recently by one of
us\cite{unpublished}, for the case of a circular orbit.
A more detailed account, including a more systematic presentation of the
regularization procedure, full details of the calculations involved
(for radial trajectories), and
generalization to a wider class of static spherically symmetric black hole
spacetimes (including, for example, the Reissner-Nordstr\"{o}m and
Schwarzschild-de Sitter spacetimes),
will be given in a forthcoming paper\cite {BO}.

Schematically, the calculation of the self-force in our approach
contains two (essentially independent) parts.
In the first part, which deals with the contribution of the individual
modes, the appropriate ordinary DE
for each mode $\ell m\omega$ of the field is solved
(numerically, in most cases), and the contribution
$F_{\alpha}^{\ell m\omega}$
of each mode to the self-force is evaluated.
(Alternatively, one may numerically solve the 1+1 partial DE in the time
domain, for each $\ell$ and $m$.)
In the second part, a certain regularization procedure is applied
to the mode sum. This part involves the analytic calculation of certain
regularization parameters, using a local perturbative analysis.
This manuscript deals with the second part, namely, the analytic
regularization scheme. The first part---the numerical determination
of $F_{\alpha}^{\ell m\omega}$---was implemented by Burko
for several scenarios\cite{Burko1,Burko2,Amaldi}.

Throughout this paper we use relativistic units (with $G=c=1$),
metric signature $({-}{+}{+}{+})$, and Schwarzschild coordinates
$t,r,\theta,\varphi$.

The total self-force on a scalar particle is composed of three parts
\cite{QW2,Qu&Wi}:
\begin{equation}\label{eq10}
F^{\rm (total)}_{\alpha}=F^{\rm (ALD)}_{\alpha}+F^{\rm (Ricci)}_{\alpha}
+F^{\rm (tail)}_{\alpha}.
\end{equation}
The first term is an Abraham-Lorentz-Dirac (ALD)-like term,
$F^{\rm (ALD)}_{\alpha}=\frac{1}{3}q^2(\dot{a}_{\alpha}-a^2 u_{\alpha})$,
where $u^{\alpha}$ and $a^{\alpha}$ are,
respectively, the four-velocity and four-acceleration of the particle,
$a^2\equiv a_{\beta}a^{\beta}$,
and an overdot represents covariant differentiation with respect to the
particle's proper time $\tau$.
The second term is a Ricci-related term,
which vanishes in the Schwarzschild case considered here.
The third term, $F^{\rm (tail)}_{\alpha}$, which
represents a non-local contribution to the self-force,
may be expressed as
\begin{equation}\label{eq20}
F^{\rm (tail)}_{\alpha}\equiv\lim_{\epsilon\to 0^+}F^{(\epsilon)}_{\alpha},
\end{equation}
where\footnote {Following \cite{QW2}, we
use here $F_{\mu}=q\phi_{,\mu}$ as the basic expression for the force
applied on a charged particle by the scalar field.}
\begin{equation}\label{eq30}
F^{(\epsilon)}_{\alpha}\equiv q^2\int_{-\infty}^{-\epsilon}
G_{,\alpha}\left[x^{\mu};x_p^{\mu}(\tau)\right] d\tau.
\end{equation}
Here, $x_p^{\mu}(\tau)$ denotes the particle's worldline,
and the integrand is evaluated at
$x^{\mu}=x_0^{\mu}\equiv x_p^{\mu}(\tau=0)$ (the self-force
evaluation point). $G$ is the retarded Green's function, satisfying
\begin{equation}\label{eq40}
\Box G(x^{\mu};{x'}^{\mu})=\frac{-4\pi}{\sqrt{-g}}\,\,
\delta^4(x^{\mu}-{x'}^{\mu}),
\end{equation}
subject to the causal initial condition, i.e., $G=0$ whenever $x^{\mu}$ lies
outside the future light cone of ${x'}^{\mu}$. [In Eq.\ (\ref{eq40}),
$\Box$ represents the covariant D'Alembertian operator.]
From the practical point of view, the challenging task is the calculation
of the tail part, which requires knowledge of the Green's function
everywhere along the particle's past worldline.
In what follows we therefore focus on the calculation of the tail term
$F^{\rm (tail)}_{\alpha}$.

We decompose $F^{(\epsilon)}_{\alpha}$ and $G(x^{\mu};{x'}^{\mu})$
into spherical harmonic $\ell$-modes,
\begin{equation}\label{eq50}
F^{(\epsilon)}_{\alpha}=\sum_{\ell=0}^{\infty}F^{\ell(\epsilon)}_{\alpha}
,\quad\quad
G=\sum_{\ell=0}^{\infty}G^{\ell}
\end{equation}
($F^{\ell(\epsilon)}_{\alpha}$ and $G^{\ell}$ are the quantities
resulting from summation over the azimuthal number $m$).
We can then write
\begin{equation}\label{eq60}
F^{\rm (tail)}_{\alpha}= \lim_{\epsilon\to 0^+}\sum_{\ell}
F^{\ell(\epsilon)}_{\alpha} =
\lim_{\epsilon\to 0^+}\sum_{\ell}\left(F^{\ell}_{\alpha}
-\delta F^{\ell(\epsilon)}_{\alpha} \right),
\end{equation}
where $F^{\ell(\epsilon)}_{\alpha}$, $\delta F^{\ell(\epsilon)}_{\alpha}$,
and $F^{\ell}_{\alpha}$ denote the force associated with the $\ell$ multipole of
the field sourced, respectively, by the interval $\tau \leq -\epsilon$,
the interval $\tau > -\epsilon$, and the entire worldline
(through the $\ell$-mode Green's function $G^{\ell}$).
The quantity $F^{\ell}_{\alpha}$
may be identified with the sum over $m,\omega$
of the contributions from all modes $\ell m\omega$ for a given $\ell$
(recall that in calculating the total field's mode $\ell m\omega$
one takes the source term to be the {\it entire} worldline).

It should be emphasized that the limit $\epsilon \to 0$ and the
sum over $\ell$ in Eq. (\ref{eq60}) should {\it not} be interchanged
(otherwise the entire contribution from $\delta F^{\ell(\epsilon)}_{\alpha}$,
which is crucial for the regularization procedure, would be lost).
Also, a clarification is required here concerning the meaning of the last
equality in Eq.\ (\ref{eq60}):
The quantity $F^{\ell(\epsilon)}_{\alpha}$ is
well defined at $r=r_0$ (the particle's location).
The situation with $F^{\ell}_{\alpha}$ and $\delta
F^{\ell(\epsilon)}_{\alpha}$ is more delicate, however. Each of these
quantities has well defined values at the limits $r\to r_0^-$ and
$r\to r_0^+$. However, these two one-sided values are
not the same. Equation (\ref{eq60})
should thus be viewed as an equation for either the limit $r\to r_0^-$
of the relevant quantities (i.e., $F^{\ell}_{\alpha}$ and
$\delta F^{\ell(\epsilon)}_{\alpha}$), or the limit $r\to r_0^+$ of these
quantities. Obviously, this equation is also valid for the {\it averaged
force}, i.e., the average of these two one-sided values.
Throughout this paper we
shall always consider the averaged force.
(Of course, the final result of the calculation, $F^{\rm
(tail)}_{\alpha}$,
having a well defined value at the evaluation point, should be the
same regardless of whether it is derived from its one-sided limit
$r\to r_0^-$, or from $r\to r_0^+$, or from their average.)

Next, we seek a simple $\epsilon$-independent function $h^{\ell}_{\alpha}$,
such that the series $\sum_{\ell}(F^{\ell}_{\alpha} - h^{\ell}_{\alpha})$
would converge.
Once such a function is found, Eq.\ (\ref{eq60}) becomes
\begin{equation}\label{eq70}
F^{\rm (tail)}_{\alpha}=\sum_{\ell}\left(F^{\ell}_{\alpha}-h^{\ell}_{\alpha}
\right) - D_{\alpha},
\end{equation}
where
\begin{equation}\label{eq80}
D_{\alpha} \equiv \lim_{\epsilon\to 0^+}\sum_{\ell} \left(\delta
F^{\ell(\epsilon)}_{\alpha} - h^{\ell}_{\alpha}\right).
\end{equation}
In principle, $h^{\ell}_{\alpha}$ can be found by investigating the
asymptotic behavior of $F^{\ell}_{\alpha}$ as $\ell \to \infty$.
It is also possible, however, to derive $h^{\ell}_{\alpha}$ from the
large-$\ell$ asymptotic behavior of $\delta F^{\ell(\epsilon)}_{\alpha}$
[the latter and $F^{\ell}_{\alpha}$ must have the same
large-$\ell$ singular asymptotic behavior (for fixed $\epsilon$),
because their difference yields a convergent sum over $\ell$].
Obviously, in order to determine $h^{\ell}_{\alpha}$ and $D_{\alpha}$
from $\delta F^{\ell(\epsilon)}_{\alpha}$,
one only needs the asymptotic behavior of $\delta F^{\ell(\epsilon)}_{\alpha}$
in the immediate neighborhood of $\epsilon=0$. This allows one
to derive $h^{\ell}_{\alpha}$ and $D_{\alpha}$
using local analytic methods, as we now describe.

To proceed, we define the (Eddington-Finkelstein) null coordinates,
$v\equiv t+r_*$ and $u\equiv t-r_*$,
where $r_*\equiv r+2M\ln(r/2M-1)$.
The Green's function at $x^{\mu}$ for a
source point at ${x'}^{\mu}$ may be expressed as \cite{BO}
\begin{equation}\label{eq100}
G^\ell = L P_{\ell}(\cos\chi) \frac{g^\ell (v,u;v',u')}{r r'}
\Theta(v-v')\Theta(u-u'),
\end{equation}
where $L\equiv \ell+1/2$,
$\Theta$ is the standard step function, $P_{\ell}$ is the
Legendre polynomial, and
\begin{equation}\label{eq110}
\cos\chi=\cos\theta \cos\theta' + \sin\theta \sin\theta'
\cos(\varphi-\varphi').
\end{equation}
The ``reduced Green's function'' ${g^\ell (v,u;v',u')}$ satisfies
\begin{equation}\label{eq120}
g_{,vu}^{\ell}+V^{\ell}(r)g^{\ell}=0,
\end{equation}
with the supplementary characteristic initial conditions
$g^\ell (v=v')=g^\ell (u=u')=1$.
The effective potential is $V^{\ell}(r)= (f/4)[\ell (\ell+1)/r^2+f'/r]$,
where $f\equiv 1-2M/r$ and a prime denotes $d/dr$.
We re-express it as $V^\ell (r)=L^2 V_0(r)+V_1(r)$,
where $V_0=f/4r^2$ and $V_1=(f/16r^2)(4rf'-1)$.

To explore the asymptotic behavior of the function $g^{\ell}$
for small spacetime intervals (and for large $\ell$),
we next apply the following perturbation analysis.
We first Taylor-expand $V^{\ell}(r)$ in the small deviation of $r$ from
$r_0$. It is convenient to take the small expansion parameter to be
$r_*-r_{*0}$. We also define
\begin{equation}\label{eq160}
\Delta_{r}\equiv 2V_{00}^{1/2}L(r_*-r_{*0}),
\end{equation}
where $V_{00}\equiv V_0 (r_0)$. Expressing $r_*-r_{*0}$ in terms of
$\Delta_{r}/L$, substituting in the Taylor expansion, and
collecting terms of the same powers in $L$ (with fixed $\Delta_{r}$),
one finds
\begin{eqnarray}\label{eq150}
V^\ell (r)&=&V_{00}\left[L^2+
L\left(f_1 \Delta_{r}\right)+ \left(f_2+f_3 \Delta_{r}^2\right)
\right]\nonumber\\&&+O(1/L),
\end{eqnarray}
where $f_1$, $f_2$, and $f_3$ are coefficients given by
\begin{eqnarray}\label{eq170}
f_1&\equiv& \frac{1}{2}V_{00}^{-3/2}\bar{V_0}=f^{-1/2}(r f'-2f),
\\
f_2&\equiv&  V_{00}^{-1}V_1 = r f'-1/4,
\\
f_3&\equiv& \frac{1}{8}V_{00}^{-2}\bar{\bar{V_0}}=
\frac{r^2}{2}\left[(f')^2/f+f''\right]+3(f-rf').
\end{eqnarray}
Here, an overbar denotes $d/dr_*$, and all quantities are evaluated at
$r=r_0$. We now expand $g^\ell$ in the form
\begin{equation}\label{eq180}
g^\ell =\sum_{k=0}^{\infty}L^{-k}g_k (\Delta_{r},\Delta_{r'},z),
\end{equation}
where $\Delta_{r'}\equiv 2V_{00}^{1/2}L(r'_*-r_{*0})$ and
\begin{equation}\label{eq200}
z\equiv 2L\left[V_{00}(v-v')(u-u')\right]^{1/2}.
\end{equation}
(The variable $z$ represents the geodesic distance, scaled by $L/r_0$,
between the Green's function evaluation and source points, to leading
order in $r-r'$.)
Substituting the expansions (\ref{eq150}) and (\ref{eq180}) in
Eq.\ (\ref{eq120}) and comparing powers of $L$, we obtain a hierarchy of
equations for the various functions $g_k$, of the form
\begin{equation}\label{eq21n}
g_{k,yx}+g_k =S_k.
\end{equation}
Here, $y\equiv V_{00}^{1/2}L(v-v')$,
$x\equiv V_{00}^{1/2}L(u-u')$,
and $S_k$ is a source term determined by $g_{k'<k}$.
In the analysis below we shall only need the terms $k=0,1,2$.
For these values of $k$, the source terms are $S_0=0$,
$S_1=-f_1\Delta_{r}g_0$, and
$S_2=-f_1\Delta_{r}g_1 -(f_2+f_3\Delta_{r}^2)g_0$.

The solution to Eq.\ (\ref{eq21n}) for $k=0,1,2$, subject to the
initial conditions
$g_k (v=v')=g_k (u=u')=\delta_{k0}$
(which conform with the original initial conditions for $g^\ell$),
is given by\cite{BO}
\begin{equation}\label{eq240}
g_0 =J_0(z),
\end{equation}
\begin{equation}\label{eq250}
g_1 =-\frac{1}{4}f_1 zJ_1(z)(\Delta_{r}+\Delta_{r'}),
\end{equation}
\begin{eqnarray}\label{eq260}
g_2 &=&-\frac{1}{6}zJ_1(z)\left[f_3
(\Delta_{r}^2+\Delta_{r}\Delta_{r'}+\Delta_{r'}^2)
                     +3f_2\right] \nonumber\\
&&+\frac{1}{96}z^2J_2(z)\left[3f_1^2(\Delta_{r}+
               \Delta_{r'})^2-8f_3\right] \nonumber\\
&&+\frac{1}{96}f_1^2 z^3J_3(z),
\end{eqnarray}
where $J_n(z)$ are the Bessel functions of the first kind.

Combining Eqs.\ (\ref{eq100}) and (\ref{eq180}), one obtains an expression
for $G^\ell$ in powers of $1/L$. One next constructs $\delta
F^{\ell(\epsilon)}_{\alpha}$ by
\begin{equation}\label{eq270}
\delta F^{\ell (\epsilon)}_{\alpha}= q^2\int_{-\epsilon}^{0^+}
G^\ell_{,\alpha}\left[x^{\mu};x_p^{\mu}(\tau)\right] d\tau
\end{equation}
(evaluated at $x^{\mu}=x_0^{\mu}$).

From this point on, we shall restrict attention to radial trajectories.
In this case, the only
non-vanishing components of the self-force are $F_r$ and $F_t$
[note that $P_{\ell}(\cos\chi)=1$].
Expanding the various $\tau$-dependent quantities involved in the
calculation (such as $r'$, $\Delta_{r'}$, and $z$) in powers of
$\tau$, replacing $\tau$ by $L^{-1} (L\tau)$, and collecting powers of $1/L$
(for fixed $L\tau$),
one obtains an expression for $\delta F^{\ell(\epsilon)}_{\alpha}$
in the form of an expansion in $1/L$,
\begin{equation}\label{eq290}
\delta F^{\ell(\epsilon)}_{\alpha}=\int_{0}^{\epsilon L}
\left[LH_{\alpha}^{(0)}+H_{\alpha}^{(1)}+
H_{\alpha}^{(2)}/L+O(L^{-2})\right] d\lambda,
\end{equation}
where $\lambda \equiv -L\tau$. The coefficients $H_{\alpha}^{(k)}$
are functions of $\lambda$ (but not of $L$ or $\tau$ otherwise),
which also depend on the parameters
$r_0$, $u^r$, $\dot u^r$, and $\ddot u^r$ (evaluated at $x_0^{\mu}$).
These functions, all being certain linear combinations of a
few terms $\lambda^i J_n(\lambda)$ (with integers $i$ and $n$),
will be given explicitly in Ref. \cite{BO}.

Considering the asymptotic form of Eq.\ (\ref{eq290}) for large $\ell$,
one finds that the regularization function $h^\ell _{\alpha}$
takes the form
\begin{equation}\label{eq300}
h^\ell _{\alpha}=A_{\alpha}L+B_{\alpha}+C_{\alpha}/L,
\end{equation}
where\footnote{
The integrals below include oscillatory terms
which do not converge as $\lambda \to \infty$.
In our calculation scheme we simply drop these terms.
This will be justified in Ref. \cite{BO}.}
\begin{eqnarray}\label{eq310}
A_{\alpha}&=&  \int_{0}^{\infty} H_{\alpha}^{(0)}(\lambda)d\lambda,
\quad\quad
B_{\alpha}= \int_{0}^{\infty} H_{\alpha}^{(1)}(\lambda)d\lambda,
\nonumber\\
C_{\alpha}&=& \int_{0}^{\infty} H_{\alpha}^{(2)}(\lambda)d\lambda.
\end{eqnarray}
Substituting Eqs. (\ref{eq290}) and (\ref{eq300}) in Eq.\ (\ref{eq80}),
one obtains
\begin{equation}\label{eq320}
D_{\alpha}=-\lim_{\epsilon\to 0^+}\sum_\ell \int_{L\epsilon}^{\infty}
\left[LH_{\alpha}^{(0)}+H_{\alpha}^{(1)}+H_{\alpha}^{(2)}/L\right]
d\lambda
\end{equation}
[the contribution from the $O(L^{-2})$ term in Eq. (\ref{eq290}) vanishes
at the limit $\epsilon \to 0$ \cite{BO}].

In conclusion, the tail part of the self-force is given by
\begin{equation}\label{eq330}
F^{\rm (tail)}_{\alpha}=\sum_{\ell}\left(F^{\ell}_{\alpha}-A_{\alpha}L
-B_{\alpha}-C_{\alpha}/L\right)-D_{\alpha}.
\end{equation}
(From the above construction of $h_{\alpha}^{\ell}$ it follows that the
sum over $\ell$ converges like $\sim\ell^{-1}$.)
The implementation of our regularization scheme thus amounts to
analytically determining the regularization parameters $A_{\alpha}$,
$B_{\alpha}$, $C_{\alpha}$, and $D_{\alpha}$, using Eqs.\ (\ref{eq310})
and (\ref{eq320}).
We point out that, although Eq.\ (\ref{eq330}) has been developed here
for a radial motion, an expression of this form is also valid for any
other trajectory. (Of course, the details of the trajectory under
consideration will enter the calculation of the regularization parameters.)

We shall now present the results obtained for the values of the above
regularization parameters in the case of a radial motion
(not necessarily geodesic). First, one finds\cite{BO}
that both parameters $A_{\alpha}$ and $C_{\alpha}$
vanish identically:
\begin{equation}\label{eq340}
A_{\alpha}=C_{\alpha}=0.
\end{equation}
In fact, it can be shown that at least the $r$ component $A_r$ vanishes
for {\it all} orbits in the Schwarzschild geometry\cite{unpublished}, and
preliminary results indicate that the same holds for $C_r$.
(Recall that we are considering here the averaged force: the
``one-sided'' values of $A_{\alpha}$ are, in general, nonzero\cite{BO}.)
The vanishing of $C_{\alpha}$ plays a crucial role in the regularization
scheme: as it turns out\cite{BO}, the parameter $D_{\alpha}$ cannot be
well-defined [specifically, the limit $\epsilon \to 0$
in Eq.\ (\ref{eq80}) diverges] unless $C_{\alpha}$ vanishes.
Thus, the result (\ref{eq340}) may be viewed as a demonstration of the
scheme's self-consistency.

On the other hand, the parameters $B_{\alpha}$ and $D_{\alpha}$ are
generically nonvanishing.
For a radial motion in Schwarzschild, a calculation
based on Eq.\ (\ref{eq310}) yields \cite{BO}
\begin{equation}\label{eq350}
B_{\alpha}^{\rm (radial)}=-\frac{q^2}{2r_0^2}\left(\delta_{\alpha}^{r}+
r_0 a_{\alpha} - \dot r u_{\alpha}\right).
\end{equation}
The calculation of $D_{\alpha}$ is more complicated.
Basically, it involves transforming in Eq.\ (\ref{eq320}) from a summation
over $\ell$ to an integration over $L$
\footnote{When performing this transformation,
one must also take into account the $O(\epsilon^2)$ contribution arising
from the difference between the sum over $\ell$ and the corresponding
integral. \cite{BO}},
followed by an evaluation of the resulting double integral in the limit
$\epsilon\to 0$. This calculation, whose details will be presented
in Ref.\ \cite{BO}, yields, in the case of radial motion,
\begin{equation}\label{eq360}
D_{\alpha}=\frac{1}{3}q^2 \left(\dot{a}_{\alpha}-a^2 u_{\alpha}\right).
\end{equation}
Remarkably, $D_{\alpha}$ is exactly the same as the ALD-like term
$F_{\alpha}^{\rm (ALD)}$.
Thus, the contribution of $D_{\alpha}$ to the tail term
{\it cancels out} the local term in the
expression for the total self-force, Eq.\ (\ref{eq10}),
leading to the simple result
\begin{equation}\label{eq380}
F^{\rm (total)}_{\alpha}=\sum_{\ell}\left(F^{\ell}_{\alpha}
-B_{\alpha}\right).
\end{equation}
Moreover, our analysis of a more general class of
non-vacuum, static, spherically symmetric
spacetimes (for which, in general, $F_{\alpha}^{\rm (Ricci)}\neq 0$)
yields\cite{BO}
\begin{equation}\label{eq385}
D_{\alpha}=F_{\alpha}^{\rm (ALD)}+F_{\alpha}^{\rm (Ricci)},
\end{equation}
implying that Eq.\ (\ref{eq380}) is valid in this more general non-vacuum
class as well.

The regularization method described above has also been implemented for the
case of a scalar particle in a uniform {\it circular} motion
around a Schwarzschild black hole\cite{unpublished}.
Here we merely present the results,
concerning the $r$-component of the regularization parameters in this case.
For circular orbits, the parameter $D_r$ (as well as $A_r$ and
$C_r$) is found to vanish.\footnote{
For circular orbits, $F_{r}^{\rm (ALD)}=0$.
Thus, Eq.\ (\ref{eq380}) is valid
for the $r$ component in circular motion as well.
From the analysis of Ref.\ \cite {Ori} (concerning the evolution rate
of $E$) it follows that in circular motion Eq.\ (\ref{eq380}) holds also
for the $t$ component (with $B_t=0$; see also \cite{Burko2}).}
The parameter $B_r$ takes the form \cite{unpublished}
\begin{equation}\label{eq390}
B_r^{\rm (circular)}=
-\frac{q^2}{2r_0^2 \gamma}
\left({2\,I_a-\frac{r_0-3M}{r_0-2M}\,I_b} \right).
\end{equation}
Here, $\gamma\equiv (1-V^2)^{-1/2}$,
$V$ is the tangential velocity with respect to a static observer, i.e.,
$V=r_0 (-g_{tt})^{-1/2}(d\varphi/dt)$ (for an equatorial orbit), and
$I_a=F(1/2,1/2;1;V^2)$ and $I_b=F(1/2,3/2;1;V^2)$
are hypergeometric functions.

A special case of both Eqs.\ (\ref{eq350}) and (\ref{eq390})
is the one corresponding to a {\em static} particle, for which we find
\begin{equation}\label{eq400}
B_r^{\rm (static)}=
-\frac{q^2}{2r_0^2}\left(\frac{r_0-M}{r_0-2M}\right).
\end{equation}

Recently, Burko numerically calculated $F^{\ell m\omega}$ for static
\cite {Burko1} and circular \cite {Burko2} orbits, and used the
above regularization scheme to calculate the self-force.
The above expressions for $B_r$ [Eqs.\ (\ref{eq390}) and (\ref{eq400})],
as well as the vanishing of $A_r$ and $C_r$,
are in excellent agreement with the large-$\ell$ limit of his
numerically-deduced $F^{\ell}$ (obtained by summing $F^{\ell m\omega}$
over $m$ and $\omega$).

To conclude, in this paper we have presented a practical prescription
for the calculation of the self-force acting on a scalar particle in the
Schwarzschild spacetime.
This prescription involves the (essentially straightforward)
numerical calculation of the mode contributions $F^{\ell}_{\alpha}$,
and the analytic derivation of four regularization parameters.
The tail part of the self-force is then given by Eq.\ (\ref{eq330}).
For radial trajectories, the {\em total} self-force takes the simple form
(\ref{eq380}), with $B_{\alpha}$ given by Eq.\ (\ref{eq350}).
Based also on other cases studied so far\cite{unpublished,BO}
one may, perhaps, conjecture that Eq.\ (\ref{eq380}) holds for {\em any}
trajectory in {\em any} spherically symmetric spacetime.

The generalization of this mode-sum regularization scheme to general
trajectories in the Schwarzschild spacetime should be straightforward,
based on the explicit expressions given above for the
Green's function, Eqs.\ (\ref{eq100}) and (\ref{eq240})--(\ref{eq260}).
In Ref.\ \cite{BO} we further extend the regularization scheme to a
more general class of static spherically symmetric spacetimes.
We expect that the generalization to the electromagnetic
self-force will be almost straightforward. The important task of
generalizing the method to the Kerr case and to the gravitational
self-force seems more challenging; yet, we believe it should also be
possible.

Finally, we note that a closely related regularization approach, which is
also based on the multipole expansion, is currently being
studied by Lousto \cite{Lousto}.

We are grateful to Lior Burko for many interesting discussions and
helpful comments. One of us (A.O.) wishes to thank Kip Thorne
for his warm hospitality during a visit in Caltech,
where this work was initiated.



\end{document}